\documentclass[conference]{IEEEtran}

\usepackage{booktabs}
\IEEEoverridecommandlockouts
\usepackage{cite}
\usepackage{amsmath,amssymb,amsfonts}
\usepackage{algorithmic}
\usepackage{graphicx}
\usepackage{textcomp}
\usepackage{xcolor}
\usepackage{hyperref}
\usepackage{tabularx}

\usepackage{xcolor}
\usepackage[most]{tcolorbox}

\usepackage{pifont}

\usepackage{booktabs}
\usepackage{array}
\usepackage{pifont}
\usepackage[table,xcdraw]{xcolor}

\usepackage{hyperref}
\usepackage{xcolor}
\usepackage{fontawesome}

\usepackage[switch]{lineno}

\usepackage[switch]{lineno}
  
\def\BibTeX{{\rm B\kern-.05em{\sc i\kern-.025em b}\kern-.08em
    T\kern-.1667em\lower.7ex\hbox{E}\kern-.125emX}}
    
\begin{document}

\title{ RE for AI in Practice:  Managing Data Annotation Requirements for AI Autonomous Driving Systems }

\author{\IEEEauthorblockN{1\textsuperscript{st} Hina Saeeda}
\IEEEauthorblockA{\textit{Dept. Computer Science and Engineering} \\
\textit{Chalmers University of Technology}\\
Gothenburg, Sweden \\
hinasa@chalmers.se}
\and
\IEEEauthorblockN{2\textsuperscript{nd} Mazen Mohamad}
\IEEEauthorblockA{\textit{RISE Research Institutes} \\
Gothenburg, Sweden \\
mazen.mohamad@ri.se}
\and
\IEEEauthorblockN{3\textsuperscript{rd}Eric Knauss}
\IEEEauthorblockA{\textit{Dept. Computer Science and Engineering} \\
\textit{University of Gothenburg}\\
eric.knauss@cse.gu.se}

\and
\IEEEauthorblockN{4\textsuperscript{th} Jennifer Horkoff}
\IEEEauthorblockA{\textit{Dept. Computer Science and Engineering} \\
\textit{Chalmers University of Technology}\\
Gothenburg, Sweden \\
jenho@chalmers.se}
\and
\IEEEauthorblockN{5\textsuperscript{th} Ali Nouri}
\IEEEauthorblockA{\textit{Volvo Cars } \\
Gothenburg, Sweden\\
ali.nouri@volvocars.com}
\and
}


\maketitle
\vspace{-0.5cm}\vspace{-0.3cm}
\begin{abstract}
High-quality data annotation requirements are crucial for the development of safe and reliable AI-enabled perception systems (AIePS) in autonomous driving. Although these requirements play a vital role in reducing bias and enhancing performance, their formulation and management remain underexplored, leading to inconsistencies, safety risks, and regulatory concerns. Our study investigates how annotation requirements are defined and used in practice, the challenges in ensuring their quality, practitioner-recommended improvements, and their impact on AIePS development and performance. We conducted $19$ semi-structured interviews with participants from six international companies and four research organisations. Our thematic analysis reveals five main key challenges: ambiguity, edge case complexity, evolving requirements, inconsistencies, and resource constraints and three main categories of best practices, including ensuring compliance with ethical standards, improving data annotation requirements guidelines, and embedded quality assurance for data annotation requirements.  We also uncover critical interrelationships between annotation requirements, annotation practices, annotated data quality, and AIePS performance and development, showing how requirement flaws propagate through the AIePS development pipeline. To the best of our knowledge, this study is the first to offer empirically grounded guidance on improving annotation requirements, offering actionable insights to enhance annotation quality, regulatory compliance, and system reliability. It also contributes to the emerging fields of Software Engineering (SE for AI) and Requirements Engineering (RE for AI) by bridging the gap between RE and AI in a timely and much-needed manner.

\end{abstract}

\begin{IEEEkeywords}
Autonomous driving, Data annotation, Data annotation requirements, AI-enabled  perception systems, Edge cases, Ethical AI, Data annotation quality, SE for AI, RE for AI
\end{IEEEkeywords}

\vspace{-0.2cm}

\section{Introduction}

AI-enabled automotive perception systems (AIePS) are central to automated driving, supporting object detection, tracking, and classification for enhanced safety and efficiency~\cite{bachute2021autonomous}. These systems underpin advanced driver assistance systems (ADAS), which offer advantages over manual driving in safety, cost, and convenience~\cite{kukkala2018advanced, khattak2021taxonomy}. Key ADAS functionalities—such as pedestrian detection, traffic sign recognition, and collision avoidance—rely heavily on sensor data from cameras, radar, lidar, and ultrasonics~\cite{chacon2015detecting, heyn2023automotive, sharma2020evaluation}.

The performance of AIePS hinges on the quality of annotated data used for training and validation~\cite{najafi2024performance, heyn2023automotive}. Inaccurate or inconsistent annotations, often caused by inadequate guidelines, human errors, or poor-quality control, can result in misclassification, unsafe behaviours, and impaired sensor fusion—ultimately compromising system reliability~\cite{chen2022road, habibullah2023requirements, yang2023uncertainties, galvao2023pedestrian, zhong2022detecting}.

A crucial yet underexplored factor in annotation quality is the role of \textit{data annotation requirements}—the standards, criteria, and instructions guiding annotation efforts~\cite{kruger2022keynote, klie2024analyzing}.  Unlike traditional software requirements that define system functionality, annotation requirements govern the preparation of training data, directly influencing the learning outcomes of AI systems. Ambiguities, missing domain rules, and insufficient quality control in these requirements can severely impact AIePS performance, particularly in safety-critical domains like autonomous driving~\cite{dey2023multi, samuktha2024framework, mohammedali2023influence}.
While RE has long addressed software requirement quality~\cite{montgomery2022empirical}, data annotation requirements remain poorly defined and lack improvement guidance~\cite{watson2023augmented, habibullah2023requirements}. This gap signals an opportunity to extend RE into AI development, a timely contribution to the emerging RE for the AI field.

To enhance the understanding and quality of data annotation requirements, we examine how they are defined and specified, their impact on annotation quality and AIePS performance, the challenges that hinder their effectiveness, and practical strategies for improvement. We explore the following research questions:

\begin{description}
\item[RQ1:] What challenges arise in ensuring high-quality data annotation requirements, and what are their underlying causes and consequences?

\item[RQ2:] What are the practitioners’ recommendations to improve data annotation requirements?

\end{description}

We conducted $(n=19)$ semi-structured interviews with $(n=20)$ participants from six international companies and four research institutes. To the best of our knowledge, this study is the first to examine data annotation requirements as a formal and evolving concern in RE for AI. We present empirically grounded insights into key challenges, best practices, and their interrelationships, offering actionable guidance to improve annotation quality and AI system development and performance. Unlike traditional system requirements, annotation requirements directly shape the training data on which AI systems learn, making their quality essential for safety and reliability.
The remainder of the paper presents background and related work, describes our research methodology, outlines key findings, discusses implications and validity threats, and concludes with recommendations for future research.

\section{Background and Related Work}

AIePS are critical for autonomous driving, supporting functions such as object detection, emergency braking, and pedestrian recognition through sensor fusion from cameras, LiDAR, and radar~\cite{heyn2023automotive, perception-systems}. Training these systems requires large volumes of annotated data~\cite{demrozi2021towards,najafi2024performance}, generated through annotation processes that convert raw sensor inputs into structured datasets. These processes are guided by \textbf{\textit{data annotation requirements, also called annotation requirements}} which define what and how to annotate, ensuring quality and consistency through mechanisms such as inter-annotator agreement~\cite{rasmussen2022challenge}.

The quality of the annotation process directly depends on the clarity and robustness of these requirements~\cite{kruger2022keynote,klie2024analyzing}. However, challenges persist due to complex sensor data, evolving or ambiguous annotation requirements, and edge case coverage\cite{fredriksson2020data}. Edge cases are rare or ambiguous scenarios outside typical data distributions that are hard to annotate consistently with standard guidelines \cite{habibullah2023requirements,heyn2023automotive}. These issues are particularly critical when labelling complex objects like pedestrians or road signs, where annotation requirements directly impact system safety and model generalisation~\cite{samuktha2024framework,borg2023ergo}.

Recent research highlights the importance of treating annotation requirements as formal, evolving artefacts. Mohammedali et al.\cite{mohammedali2023influence} demonstrate how annotation conditions influence model performance, while Heyn et al.\cite{heyn2023automotive} highlight data specification gaps in the Swedish automotive sector. Liu et al.\cite{L310509812} and Beck et al.\cite{beck2023quality} focus on annotation quality but do not address how annotation requirements should be defined or improved.

Our study empirically explores annotation requirements as a formal aspect of RE for AI, highlighting their role in guiding annotation processes and impacting the development and reliability of AIePS.

\section{Research Methodology}
\label{methodology}

\begin{figure}[htbp]  
    \centering
    \includegraphics[width=0.5\textwidth]{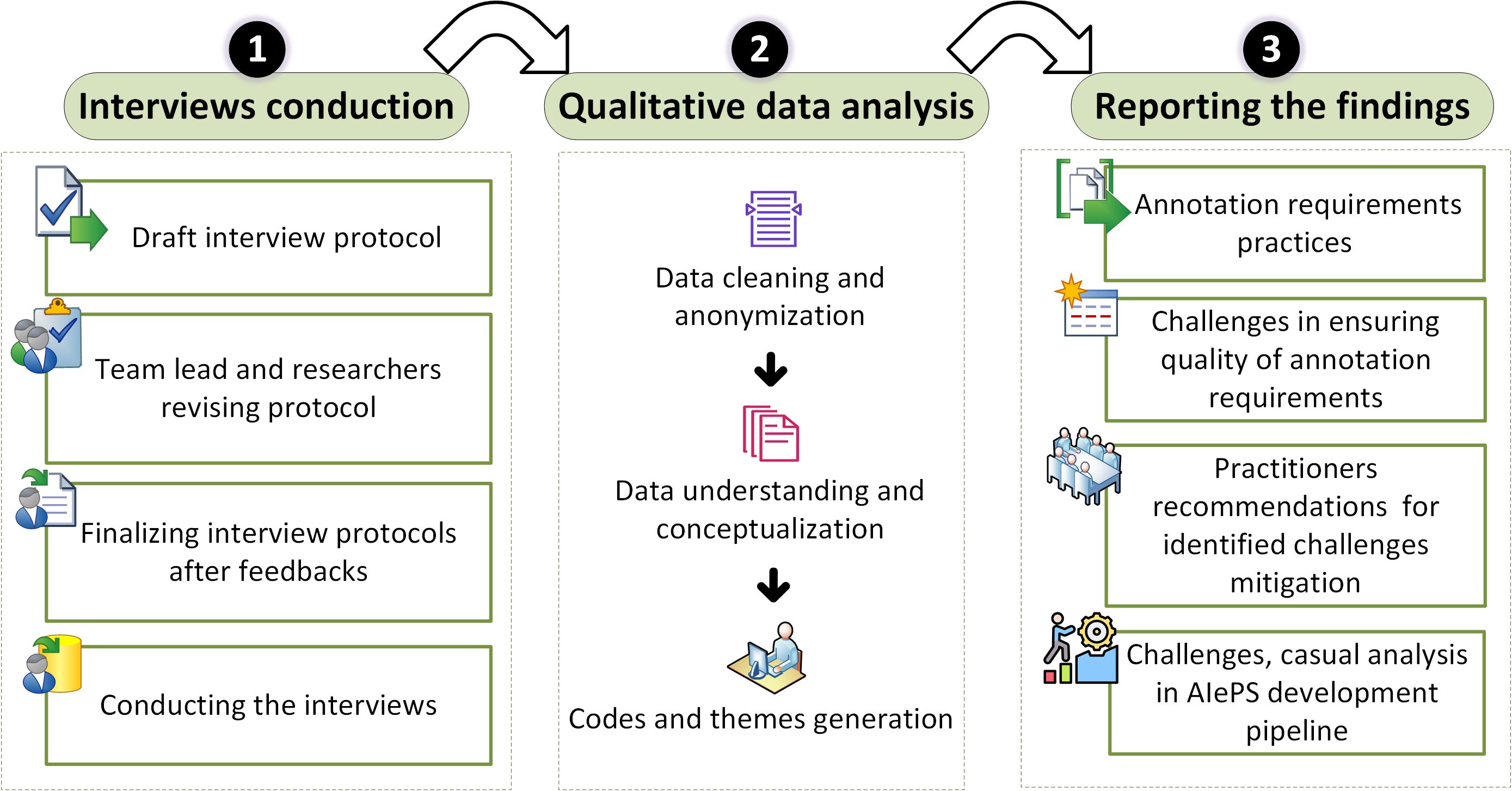} 
    \caption{ Research method followed. }
   \label{fig:method}
\end{figure}

We conducted a qualitative study, following ACM SIGSOFT Empirical Standards \cite{ralph2020empirical}, to explore definitions, practices, challenges, and recommendations for data annotation requirements in AIePS as steps detailed in Fig. \ref{fig:method}.\vspace{-0.2cm}

\textbf{Research Context} This study is part of Project X, which aims to enhance academia-industry collaboration by developing concepts, models, and techniques for effective RE in safe AIePS.
We conducted $19$ semi-structured online interviews with $20$ participants from industry and academia between September $2024$ and February $2025$, each lasting $120$ minutes~\cite{runeson2009guidelines}.  ID16 participated in two separate interviews (labeled 16A and 16B), each lasting four hours. In contrast, participants ID18 and ID19 took part in a single focus group session, counted as one interview (see Table \ref{tab:expert_table}). In total, this amounts to over 50 hours of rich transcripts.
 Using purposeful sampling~\cite{ralph2020empirical}, we selected participants based on their roles, organisational context, and experience with data annotation or AI model development in autonomous driving. Interviews were recorded, transcribed, manually corrected, and concluded with participant feedback. This approach ensured diverse and in-depth insights into annotation requirements, associated challenges, and their impact on AIePS performance.

\textbf{Companies Context} To capture diverse perspectives, we conducted fourteen interviews with representatives from six companies engaged in automotive and perception system development. We included the entire supply chain:- These included one OEM, three Tier-1 suppliers, two Tier-2 suppliers, a perception training company, and a code analysis firm, all based in two European countries and the UK. In the automotive context, OEMs produce complete vehicles (e.g., Volvo, Tesla). Tier-1 suppliers deliver key systems, such as perception sensors, directly to OEMs, while Tier-2 suppliers provide components, including embedded software, to Tier-1s.

Additionally, five interviewees from four research institutes, including a state-owned European institute and leading universities from Europe and the UK, helped bridge theory and practice across AI, data annotation, ethics, and regulation. Thematic saturation was reached by interview 19, with no new codes or themes emerging during the final three interviews (ID17, ID18, ID19).

\textbf{Participants} As shown in Table \ref{tab:expert_table} we interviewed participants based on their organisation type, professional role, and experience with data annotation or AI model development for autonomous driving. Targeted experts included leading contributors to AIePS, such as safety standard developers, founders of relevant companies, and the chief investigator of top-tier research labs, ensuring deep domain expertise. This is a niche domain where very few but globally recognised—organisations and experts are active. We deliberately reached out to these leading actors and succeeded in including the most relevant and established experts in this emerging field.

\begin{table}[!ht]
\vspace{-0.3cm}
\centering
\scriptsize
\caption{Categorisation of interviewees based on company type and specialisation, including their roles and years of experience}
    \label{tab:expert_table}
\begin{tabularx}{\linewidth}{llll}
\toprule

        \textbf{Company ID: Type} & \textbf{Focus Area} & \textbf{Expert (Years of Experience)} \\
        \midrule
        A: Tier 2 & Data Annotation & ID1: Annotation Expert (6) \\
                            && ID2: Perception Expert (10) \\
                            && ID3: Quality  Expert (9) \\
                            \midrule
        B: Tier 1 & Safety Software & ID4: Machine Learning Expert (7) \\
                             &&      ID5: Data Scientist (11) \\
                             \midrule
         C: University & Research & ID6: Senior Researcher (9) \\
         \midrule
         D: University & Research & ID7: Researcher (8) \\
         \midrule
         E: University & Research & ID8: Researcher (5) \\
         \midrule
          F: Research Institute & Research & ID9: Researcher (10)\\
                                     &&   ID10: Researcher (20) \\
                                     \midrule
        G: OEM & Automotive & ID11: Machine Learning Expert (5) \\
                           && ID12: V\&V Expert (5) \\
                          &&  ID13: Data Engineer (4) \\
                          &&  ID14: Researcher (3) \\
                          \midrule
         H: Tier 1 & Safety Software & ID15: Researcher (10) \\
         \midrule
         I: Tier 2 & Quality Assurance & ID16 A: Quality  Expert (18) \\
         I: Tier 2 & Quality Assurance & ID16 B: Quality  
         Expert (18) \\
         \midrule
         J: Tier 1 & Digital Solution & ID17: Head of Research (17)\\
                                  &&   ID18,19: Research Engineer (1,2)   \\

        \bottomrule
    \end{tabularx}
    
\end{table}

\textbf{Interview Preparation}: Before the interviews, we completed the following preparatory tasks: We developed our interview guide~\cite{ralph2020empirical} through an iterative process with industrial experts in data annotation and its requirements, ensuring open-ended, neutral, and clear questions. To enhance validity and reliability, we pre-tested it with two industry partners—representing annotation requirements (Tier 1) and production (Tier 2)—whose feedback helped refine the guide’s clarity, structure, and duration.
(For details, see: \href{https://dataverse.harvard.edu/file.xhtml?fileId=10905126&version=3.0}{\textcolor{blue}{→ link: Designed Interview Guide}}.)

\textbf{Data Analysis:} We applied a six-phase thematic analysis process outlined by Cruzes et al., \cite{runeson2009guidelines}. First, researchers familiarised themselves with the transcripts and created a deductive codebook based on interview questions~\cite{ralph2020empirical} (see: \href{https://dataverse.harvard.edu/file.xhtml?fileId=10905422&version=3.0}{\textcolor{blue}{→ link: Code Book}}). In the initial coding phase, deductive codes were applied, with inductive codes added as new insights emerged. Two researchers independently validated the coding, achieving strong inter-coder agreement $($Cohen’s kappa $= 0.8)$. Themes and sub-themes were then developed and aligned with the predefined research questions. The team collaboratively refined sub-themes by merging related codes (see: \href{https://dataverse.harvard.edu/file.xhtml?fileId=10905602&version=3.0}{\textcolor{blue}{→ link:  Merging Codes, Sub-themes and Themes }}). Finally, five researchers collaboratively conducted and reviewed the analysis using Excel, OneDrive, and Miro to ensure structured and transparent reporting. The complete thematic analysis steps are shared at  Harvard Dataverse  ( \href{https://dataverse.harvard.edu/file.xhtml?fileId=11792291&version=4.0}{\textcolor{blue}{→ link: Complete Thematic Analysis  }}). Note: (Please download the MS Excel Spreadsheet for review).


\vspace{-0.3cm}

\section{Results}

\textbf{Conceptual Anchor}  Experts emphasised that clarity in data annotation requirements leads to consistency and high-quality annotations (ID2), while vague or evolving annotation requirements cause rework and inconsistency (ID11). Poorly specified data annotation  requirements negatively impact model performance (ID18). These insights establish a causal link between annotation requirements, data quality, and system performance, informing our analysis of current practices, challenges, and recommendations.

\begin{figure}[htbp]  
    \centering
   \includegraphics[width=0.5\textwidth]{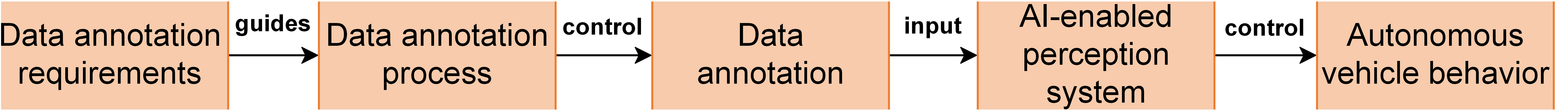} 
    \caption{ Causal relationship of annotation requirements in  AIePS development Pipeline.}
   \label{fig:Key}
\end{figure}
\vspace{-0.3cm}
   Based on this causal relationship (see Fig.~\ref{fig:Key}), we present current industry annotation practices, challenges, and expert-informed recommendations to improve requirements practices in AIePS.

\textbf{Current Practices in AIePS Requirement Management}
To contextualise the identified challenges, we first examined how annotation requirements are currently defined, updated, and governed in practice. OEMs, Tier $1$, and Tier $2$ companies adopt structured yet evolving approaches for managing annotation requirements.
These typically involve collaborative definition with stakeholders (e.g., annotators, domain experts), formalisation into traceable, version-controlled guidelines, and continuous refinement via feedback loops. Standards such as ISO/IEC $5259$, IEEE P2801, ISO $26262$, and SAE J3016 are often used in conjunction with in-house protocols, focusing on clarity, inter-annotator agreement (IAA), and safety. Teams also break down abstract legal or client-level goals into actionable instructions, particularly for edge cases. This integrated approach aligns annotation practices with technical, regulatory, and performance objectives. (For details, see: \href{https://dataverse.harvard.edu/file.xhtml?fileId=11732074&version=3.0}{\textcolor{blue}{→ link: Current Data Annotation Requirements Practices in AIePS}}.)
The next section outlines the key challenges practitioners face when implementing and evolving these data annotation requirements practices.

\subsection{\textbf{RQ1: Challenges in Ensuring High-Quality Data Annotation Requirements, their Causes and Consequences}}

\textbf{(C1) Edge Case Coverage Gaps }Edge case annotation is challenged by ambiguous, rare, and complex scenarios that resist concise standardisation.
In line with our earlier definition (Section II, Lines 107–112), we treat edge cases as rare or ambiguous scenarios outside typical data distributions that are difficult to annotate consistently with standard guidelines.
As one interviewee remarked, “You cannot have very concise guidelines and, at the same time, cover all the details [scenarios, edge cases]. It is contradictory.” (ID5). These issues are compounded by limited expertise, inadequate training, and a lack of real-world data. Lacking deep contextual knowledge makes it challenging to anticipate real deployment conditions, leading to system failures, rework, and safety or legal risks. Sub-themes C1.1 and C1.2 emphasise the importance of precisely defining and annotating edge cases to maintain annotation quality and AIePS reliability.

\textbf{(C1.1) Complexity in Annotating Edge Cases }
Annotating edge cases involves both technical and human challenges, such as unclear or evolving guidelines, annotator bias, lack of standardisation, limited training, and budget constraints. These factors hinder consistency in labelling rare yet critical events. As one participant noted, \textit{“If guidelines state to mark only visible parts, annotators may interpret this differently, introducing bias.”} (ID8). Despite using evaluation metrics such as IRR and Cohen’s Kappa, measuring agreement remains challenging due to the subjectivity of edge cases and the lack of a definitive ground truth. This complexity can lead to unreliable model performance, costly rework of annotations, and regulatory compliance risks.

\textbf{(C1.2) Challenges in Defining Requirements for Edge Cases }
Defining requirements for edge cases is inherently difficult due to their ambiguity, rarity, and context dependence. These scenarios resist standardisation, often leading to evolving, underspecified, and reactive guidelines. As one interviewee noted, \textit{“Edge cases defy standardisation [guidelines]—the more you try to pin them down, the more [coverage for] exceptions you uncover.”} (ID16). Teams frequently revise requirements after encountering gaps during annotation, resulting in dataset inconsistencies and rework. Vague definitions—such as uncertainty about whether a construction worker is a pedestrian—create annotation blind spots, elevate error rates, and compromise model reliability. Overall, the lack of proactive and well-defined requirements weakens annotation quality and reduces AIePS robustness.

\textbf{(C2) Ambiguity in Data Annotation Requirements}
Ambiguous data annotation requirements emerge as a persistent challenge that weakens annotation consistency and AIePS performance. Ambiguity stems from unclear or overly complex guidelines and the inherent subjectivity of real-world data. As one participant noted, \textit{“The challenge is that it's impossible to define unambiguous annotation requirements—images are inherently ambiguous.”} (ID16). This is further intensified by inconsistent interpretations, lack of standard definitions, and limited domain-specific training for annotators. Regulatory vagueness, such as unclear GDPR requirements, adds additional complexity (ID9). These issues lead to biased annotations, elevated error rates, and resource inefficiencies. Ambiguous requirements also hinder evaluation, making it difficult to measure annotator agreement or ensure quality. Collectively, these problems compromise traceability, increase misinterpretation risks, and diminish the reliability of both annotated data and AI system outputs.

\textbf{(C3) Challenges in Evolving Data Annotation Requirements }
Data annotation requirements are dynamic, evolving in response to regulatory shifts, technological advances, new edge cases, and changing stakeholder needs. While such adaptability allows for flexibility, it also creates challenges in maintaining clarity, consistency, and efficiency. As one interviewee observed, \textit{“Data annotation is a relatively new field... requirements are still evolving.”} (ID15). In practice, evolution is often reactive rather than proactive, lacking structured mechanisms for systematic updates. Annotation teams must quickly adapt to changes—often without updated documentation or proper training—leading to rework, inconsistencies, and operational burdens. For example, \textit{“We need to add a new property… and send annotated batches back for correction.”} (ID2). These informal updates result in guideline misinterpretation and inconsistent annotations, especially when requirements change mid-project. In fast-changing domains like autonomous driving, evolving road environments further exacerbate the issue. The lack of tool updates and reliance on outdated instructions compound these problems, ultimately affecting annotation quality and model reliability.

\textbf{(C4) Inconsistencies in Data Annotation Requirements}
 These inconsistencies arise from ad hoc requirement definitions, lack of standardisation across teams and vendors, vague or conflicting instructions, and weak communication channels. As one participant noted, \textit{“Currently, defining data annotation requirements is mostly an ad hoc process… it evolves alongside data collection.”} (ID15). Annotators often struggle with unclear purposes, as highlighted by \textit{“Annotators sometimes do not understand why they are labelling certain things.”} (ID4). Divergent interpretations between annotators and supervisors, such as \textit{“A supervisor may modify an annotation… due to a different interpretation.”} (ID3), further reduce consistency.  These issues are compounded by gaps in domain-specific training, absent version control, and inconsistent stakeholder expectations. The result is annotation variability, inefficiencies due to repeated rework, and degraded model reliability, especially in safety-critical edge cases. As one interviewee observed, \textit{“Quality inconsistencies manifest as dramatic variation across annotators.”} (ID2). Ultimately, inconsistently defined or interpreted requirements undermine traceability, generalisability, and alignment across AIePS development.

\textbf{(C5) Resource Limitations }
Resource constraints significantly influence how annotation requirements are developed and implemented in AIePS. The high cost and labour demands of data annotation lead to trade-offs that compromise quality. Underinvestment in planning and infrastructure results in rushed or reactive requirement updates, which cause ambiguities and inconsistencies (ID1, ID2, ID3). As one case revealed, skipping early planning to save costs led to poor edge case coverage later (ID15). These limitations affect the entire pipeline, increasing rework, introducing errors into model training, and undermining system reliability (ID5, ID6, ID18). Four key areas illustrate these challenges as follows: 

\textbf{(C5.1) Strict Budgets}
Budget constraints pose systemic challenges throughout the annotation pipeline, affecting the quality of personnel, tools, and quality control. Limited funding often forces reliance on undertrained annotators, leading to elevated error rates, especially in complex tasks—\textit{“Annotation is low-paid work... we can’t attract domain experts.”} (ID5). Quality assurance efforts are also reduced, such as halving manual review rates, which led to misclassifications in unchecked data (ID3, ID6). Inadequate tooling due to cost limitations forces manual workarounds, causing inconsistency—e.g., open-source tools failing to support 3D or LiDAR tasks (ID17, ID9). Some teams reduce dataset size to maintain quality, but this risks coverage gaps (ID18). Outsourcing to cut costs further compromises oversight and often results in expensive rework—\textit{“Fixing budget-driven annotation errors cost $3 \mbox{$\times$} $ more than proper initial labelling.”} (ID5).

\textbf{(C5.2) Limited Workforce and Scalability}
Workforce limitations critically hinder both annotation quality and scalability. Low pay and poor career prospects deter domain experts, forcing reliance on general-purpose workers who require extensive training—\textit{“We rely on non-specialists who need extensive training.”} (ID3). Custom onboarding for each project is time-consuming and not scalable (ID1), and despite acknowledging potential software-based training solutions, implementation is rare (ID16). High staff turnover leads to frequent retraining and wasted resources (ID5, ID18), causing quality inconsistencies and higher error rates, especially in complex or edge-case scenarios (ID2, ID8). Edge cases are often skipped to meet quotas, reducing coverage (ID9, ID12). Workforce constraints also limit throughput—some teams could annotate only $20\%$ of collected data—and scaling up often introduces new inconsistencies (ID18, ID2).

\textbf{(C5.3) Time Constraints Compromising Accuracy } Tight deadlines force trade-offs between speed and accuracy, leading to reduced annotation quality. As participants noted, \textit{“When you want to go fast, you will not be really accurate in your annotation.”} (ID7). Pressures to meet quotas or reduce costs often result in rushed work, skipped edge cases (ID12), and reduced quality control, such as lowering manual checks from $10\%$ to $5\%$ (ID3). These compromises lead to missed critical annotations (e.g., occluded pedestrians) and degrade AIePS performance due to biased or incomplete training data (ID6). In the long run, these shortcuts create significant costs—\textit{“Saved annotation days cost us months in model debugging.”} (ID18)—and rework often demands three times the original effort (ID5).

\textbf{(C5.4) Limitations in Annotation Tools and Technology }
Inadequate annotation tools significantly undermine annotation quality and scalability. Many tools lack essential features for handling high-resolution data or rare scenarios—\textit{“They lack presets for rare scenarios (e.g., occluded pedestrians at night).”} (ID12). Tool-related issues include poor version control, manual updates, and lack of collaborative capabilities, leading to inefficiencies and data loss—\textit{“Version conflicts destroyed $15\%$ of our 3D point cloud annotations.”} (ID17). Teams resort to inconsistent workarounds (ID9), and limited tool functionality restricts scalability—only a fraction of collected data can be processed (ID18). Missing audit trails and technical limitations further degrade downstream AIePS  performance, particularly in safety-critical applications—\textit{“Missing 3D annotation features led to drivable-area errors in AV testing.”} (ID6). Overall, tooling constraints amplify resource pressures and hinder the implementation of high-quality annotation requirements.

\begin{figure}[htbp]  
    \centering
   \includegraphics[width=.5\textwidth]{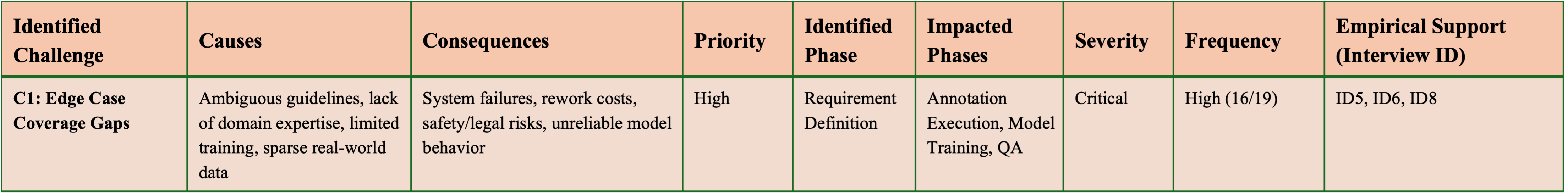} 
    \caption{ A snapshot of the publicly available challenges causal analysis dataset hosted on Harvard Dataverse. The full final repository can be found via \href{https://dataverse.harvard.edu/file.xhtml?fileId=11737896&version=3.0}{\textcolor{blue}{→ link: challenges mapped to their causes and consequences}}}
   \label{fig:ICH}
\end{figure}

\textbf{Challenges, Causal Analysis, and Criticality Summary} The example cause–and–effect analysis in Fig.~\ref{fig:ICH} reveals a hierarchy of criticality among challenges in the AIePS development pipeline. 

\vspace{-10pt}

 The most pressing concerns—classified as high-priority, high-frequency, and high-severity—include Edge Case Coverage Gaps (C1), Ambiguity in Annotation Requirements (C2), and systemic Resource Limitations (C5). These challenges, reported in over $15$ out of $19$ interviews, stem from root causes such as ambiguous or evolving guidelines, lack of domain expertise, inadequate tooling, and budget constraints. Their consequences are severe: failures in detecting rare but critical scenarios, annotation inconsistency, rework cycles, and reduced model safety and generalisation. Sub-challenges like C1.2 (Defining Requirements for Edge Cases) and C5.4 (Limitations in Annotation Tools and Technology) also received critical severity ratings due to their direct impact on safety and model performance. In contrast, moderate-priority challenges like C3 (Evolving Requirements) and C4 (Inconsistent Guidelines) still carry high frequency and affect annotation consistency, but are less likely to result in catastrophic failures if delayed. Finally, lower-priority or moderate-severity challenges, such as workforce scalability (C5.2) and time constraints (C5.3), while common, tend to impact throughput and coverage rather than safety or compliance directly. Understanding this layered challenge landscape enables a phased and risk-informed response strategy to improve annotation quality in safety-critical AIePS. 
\vspace{-0.2cm}

\begin{tcolorbox}[
  enhanced,
  colback=gray!5!white,      
  colframe=gray!60!black,    
  coltitle=white,            
  title=\textbf{Key Takeaways from RQ1},
  fonttitle=\bfseries\large,
  sharp corners,
  boxrule=0.6pt,
  drop shadow southeast,
  left=0.5mm,                
  right=2mm, top=1mm, bottom=1mm,
  before skip=10pt, after skip=10pt,
  colbacktitle=gray!60!black 
  ]

\vspace{0.3em}
\begin{itemize}[]
    \item Every annotation error ripples downstream.
    \item Structured mechanisms for annotation requirement definition are lacking.
    \item Edge cases are the Achilles' heel of annotation quality — their rarity, ambiguity, and subjectivity expose the majority of data annotation requirement limitations.
\end{itemize}

\end{tcolorbox}
\vspace{-0.3
cm}

\subsection{\textbf{RQ2: Practitioners’ recommendations to improve data annotation requirements}} \vspace{-0.2cm}
\textbf{(R1) Ensuring Compliance with Ethical Standards } Annotation requirements should align with regulatory standards such as the EU AI Act, GDPR, ISO/IEC $24027$ and ISO $26262$. These standards promote fairness, transparency, and accountability by guiding data governance, bias mitigation, and clear documentation. As one expert noted, \textit{“The AI Act emphasises data governance, pipeline design, and documentation of data assumptions.”} (ID8). Embedding ethical and legal considerations early in requirement definition reduces ambiguity, minimizes rework, and ensures consistent interpretation. In high-stakes applications, such alignment safeguards legal compliance, system reliability, and public trust.

\begin{tcolorbox}[
enhanced,
boxrule=0.3pt,
colback=white,
colframe=black!50,
left=2pt,
right=2pt,
top=2pt,
bottom=2pt,
sharp corners,
before skip=4pt,
after skip=4pt
]
\textit{$\triangleright$
R1 mitigates challenges C2, C3  and C4  by grounding annotation requirements in well-defined regulatory and ethical frameworks, thereby reducing interpretive uncertainty, guiding consistent updates, and supporting harmonised implementation across workflows.}
\end{tcolorbox}

\textbf{(R1.1) Privacy Protection and Legal Compliance in Data Annotation }
Protecting privacy and ensuring legal compliance is essential when annotating sensitive data in AIePS, such as imagery, location, or behavioural cues. This requires applying privacy-by-design principles, anonymisation methods, and strict access controls from the start. Early legal involvement helps define data-sharing agreements and align annotation processes with regulations like GDPR and local laws. As one practitioner emphasised, \textit{“If legal expectations are clear earlier, the annotation requirements won’t have to change later.”} (ID9). Proactive coordination with legal teams ensures stable, auditable requirements, avoiding costly corrections and supporting the robustness of AI models in regulated, high-risk domains.

\begin{tcolorbox}[
enhanced,
boxrule=0.3pt,
colback=white,
colframe=black!50,
left=2pt,
right=2pt,
top=2pt,
bottom=2pt,
sharp corners,
before skip=4pt,
after skip=4pt
]
\textit{$\triangleright$
R1.1 mitigates challenges C2, C3, C4, C5, and C1.2  by embedding legal safeguards and governance mechanisms early in the process, reducing misinterpretation, rework, and uncertainty around privacy-critical annotations.}
\end{tcolorbox}

\textbf{(R1.2) Defining Safety-Centric Data Annotation Requirements }
Safety-centric annotation requirements are vital to ensure that AIePS perform reliably in real-world and high-risk scenarios. Mislabeling vulnerable road users or inconsistently annotating occluded objects can lead to hazardous system behavior, such as missed detections or unsafe decisions. Critically, failing to annotate rare but important situations—like emergency vehicles or construction zones—can reduce model generalization and violate safety standards $($e.g., ISO $26262)$. As one practitioner noted, \textit{“We start by identifying data issues that could pose unacceptable safety risks.”} (ID6). Early risk assessment and context-aware annotation help distinguish tolerable from critical errors, ensuring safety is embedded throughout the annotation process.

\begin{tcolorbox}[
enhanced,
boxrule=0.3pt,
colback=white,
colframe=black!50,
left=2pt,
right=2pt,
top=2pt,
bottom=2pt,
sharp corners,
before skip=4pt,
after skip=4pt
]
\textit{$\triangleright$
R1.2 mitigates challenges C1, C1.2, C2, and C4 by grounding annotation requirements in real-world safety contexts, ensuring high-impact scenarios are adequately captured, and enforcing consistent definitions for critical cases.}
\end{tcolorbox}

\textbf{(R1.3) Ethical AI Development }
Ethical development of AIePS relies heavily on well-defined, fair, and bias-aware annotation requirements. Inaccurate or biased annotations can distort edge case representation and lead to unfair or unsafe AI behavior. To prevent this, ethical principles—such as fairness, transparency, and privacy—should be embedded from the start into annotation guidelines. As one expert emphasized, \textit{“Fairness should be a system requirement, like safety.”} (ID16). This involves aligning with frameworks like GDPR and ISO/IEC $24027$ and continuously validating ethical standards throughout development. Treating ethical requirements as core system criteria helps ensure that AI systems make responsible, context-sensitive decisions, especially in safety-critical domains.

\begin{tcolorbox}[
enhanced,
boxrule=0.3pt,
colback=white,
colframe=black!50,
left=2pt,
right=2pt,
top=2pt,
bottom=2pt,
sharp corners,
before skip=4pt,
after skip=4pt
]
\textit{$\triangleright$
R1.3 mitigates challenges C2, C4, and C5.1  by embedding fairness and bias-awareness into annotation requirements, reducing misinterpretations, supporting consistency, and preemptively addressing quality risks within limited resource constraints.}
\end{tcolorbox}

\textbf{(R2) Improving Data Annotation Requirements Guidelines }
Clear and well-structured annotation requirements are crucial for reducing errors and ensuring consistent, high-quality annotations in AIePS. As tasks grow more complex—like multi-object tracking or behavioral prediction—the risk of misinterpretation and poor model generalization increases. Practitioners warn that overly complex requirements lead to inaccuracies, even with good data—\textit{“More complexity increases inaccuracies, even with perfect data.”} (ID2). To address this, experts recommend simplifying guidelines, enhancing annotator training, integrating feedback loops, and using supportive tools to reduce cognitive burden and subjectivity during annotation.

\begin{tcolorbox}[
enhanced,
boxrule=0.3pt,
colback=white,
colframe=black!50,
left=2pt,
right=2pt,
top=2pt,
bottom=2pt,
sharp corners,
before skip=4pt,
after skip=4pt
]
\textit{$\triangleright$
R2 mitigates challenges C2, C4, and C5.2 by reducing annotator confusion, lowering error rates, and enabling more scalable and accurate annotation processes.}
\end{tcolorbox}

\textbf{(R2.1) Iterative Work with Guidelines (Plan–Do–Check–Act) }
Applying the Plan–Do–Check–Act (PDCA) cycle to data annotation fosters continuous improvement by embedding structured feedback loops and regular validation. This iterative approach ensures annotation guidelines remain aligned with evolving real-world complexities. Experts recommend using AI-assisted pre-labelling, automated quality checks, and real-time annotator feedback to streamline processes and reduce ambiguity. Visual aids, tiered reviews, and collaborative refinement sessions further support clarity and consistency. As one participant emphasized, \textit{“You need validation cycles—review and revise your requirements based on what annotators find difficult.”} (ID9). Iterative refinement promotes long-term scalability and annotation accuracy.

\begin{tcolorbox}[
enhanced,
boxrule=0.3pt,
colback=white,
colframe=black!50,
left=2pt,
right=2pt,
top=2pt,
bottom=2pt,
sharp corners,
before skip=4pt,
after skip=4pt
]
\textit{$\triangleright$
R2.1 mitigates challenges B2 (Ambiguity), B3 (Evolving Requirements), and B4 (Inconsistencies) by enabling adaptive, evidence-driven refinement of annotation requirements, improving clarity, and maintaining alignment with real-world complexities.}
\end{tcolorbox}

\textbf{(R2.2) Iterative Edge Case Development }
Edge case annotation remains challenging due to the rarity, ambiguity, and critical safety implications of such scenarios. To address this, practitioners advocate for an iterative, expert-driven process that evolves over time through annotator feedback, expert validation, and exposure to new situations. As one expert noted, \textit{“Annotators should consult domain experts for unclear cases. Edge cases should be documented and refined over time.”} (ID12). Synthetic data plays a key role in simulating rare events not present in real-world datasets (ID7). To ensure consistency, experts recommend visual aids, clear misclassification rules, and standardized QA with multi-level review protocols, especially in safety-critical domains.

\begin{tcolorbox}[
enhanced,
boxrule=0.3pt,
colback=white,
colframe=black!50,
left=2pt,
right=2pt,
top=2pt,
bottom=2pt,
sharp corners,
before skip=4pt,
after skip=4pt
]
\textit{$\triangleright$
R2.2 mitigates challenges C1.2, C2, and C4  by refining rare scenario coverage, reducing annotation errors through expert review, and improving consistency via iterative feedback and synthetic augmentation.}
\end{tcolorbox}

\textbf{(R2.3) Develop Quality Criteria and Optimise Guidelines }
Defining clear quality criteria and refining annotation guidelines are key to ensuring consistency, precision, and adaptability in data annotation workflows. Practitioners recommend using atomic labels, standardised terminology, and structured training to improve clarity. Simplifying complex instructions—using visuals and tooltips—helps reduce errors. Real-time validation tools and feedback loops allow continuous refinement, as one expert noted: \textit{“We test guidelines with annotators, refine them based on feedback, and improve clarity.”} (ID16). A/B testing and small-scale trials ensure guidelines are effective before full deployment. 
\begin{tcolorbox}[
enhanced,
boxrule=0.3pt,
colback=white,
colframe=black!50,
left=2pt,
right=2pt,
top=2pt,
bottom=2pt,
sharp corners,
before skip=4pt,
after skip=4pt
]
\textit{$\triangleright$
R2.3 mitigates challenges C2, C3, and C4  by establishing clear quality benchmarks, simplifying instructional complexity, and incorporating structured feedback loops to adapt guidelines over time.}
\end{tcolorbox}

\textbf{(R2.4) Context and Domain-Specific Data Annotation Requirements Guidelines}
Customizing annotation guidelines to the domain and context enhances accuracy and reduces ambiguity. Incorporating domain expertise, real-world examples, and visual aids helps annotators better understand complex scenarios. As one expert noted, \textit{“Domain experts improve annotation accuracy and process quality.”} (ID13). Annotators benefit from clear explanations, not just direct instructions—adding context improves understanding (ID4). Embedding tooltips, examples, and interactive exercises within annotation tools operationalizes these best practices. Feedback loops further refine guidelines to reflect real-world complexities. This approach supports clearer, more consistent annotations, particularly in specialized or high-stakes domains.

\begin{tcolorbox}[
enhanced,
boxrule=0.3pt,
colback=white,
colframe=black!50,
left=2pt,
right=2pt,
top=2pt,
bottom=2pt,
sharp corners,
before skip=4pt,
after skip=4pt
]
\textit{$\triangleright$
R2.4 mitigates challenges C2, C4, and C5.2  by embedding contextual understanding into annotation requirements, refining clarity through examples, and aligning guidelines with real-world scenarios.}
\end{tcolorbox}

\textbf{(R2.5) Rely on Automation as Support }
Integrating automation into data annotation processes improves scalability, efficiency, and consistency, particularly for repetitive or low-risk tasks. However, automation should complement—not replace—human annotators, especially in handling complex or safety-critical edge cases. As one expert noted, \textit{“Automation should assist, not replace, human annotators.”} (ID12). Pseudo-annotations from foundation models can be refined by humans (ID15), but success depends on clear guidelines, validation steps, and confidence thresholds to align automated outputs with domain expectations. This balanced approach enables scalable, reliable annotation pipelines while preserving quality through human oversight.

\begin{tcolorbox}[
enhanced,
boxrule=0.3pt,
colback=white,
colframe=black!50,
left=2pt,
right=2pt,
top=2pt,
bottom=2pt,
sharp corners,
before skip=4pt,
after skip=4pt
]
\textit{$\triangleright$
R2.5 mitigates challenges C4, C5, and C1.2  by supporting scalability through automation, while ensuring human oversight addresses complexity and edge case variability.}
\end{tcolorbox}

\textbf{(R2.6) Establishing Multidisciplinary Teams }
Creating multidisciplinary teams is essential for developing accurate, context-aware, and well-aligned data annotation requirements. Involving domain experts, data scientists, legal professionals, and annotation coordinators ensures a comprehensive understanding and reduces misalignment between technical goals and annotation practices. As one expert noted, \textit{“Cross-functional teams, including data scientists and annotation coordinators, ensure smooth coordination.”} (ID3). This collaboration supports clear requirement definition, early detection of ambiguities, and compliance with regulatory and ethical standards. Structured communication and iterative reviews help guidelines evolve effectively, reduce inconsistencies, and improve edge case handling.

\begin{tcolorbox}[
enhanced,
boxrule=0.3pt,
colback=white,
colframe=black!50,
left=2pt,
right=2pt,
top=2pt,
bottom=2pt,
sharp corners,
before skip=4pt,
after skip=4pt
]
\textit{$\triangleright$
R2.6 mitigates challenges C2, C4, C5, and C1.2  by facilitating alignment across roles, improving guideline clarity, and ensuring annotation requirements are contextually and technically grounded.}
\end{tcolorbox}

\textbf{(R2.7) Cross-Functional Collaboration in Requirement Definition }
Collaboratively defining annotation requirements with input from annotators, domain experts, tool developers, and project managers ensures that guidelines are both technically feasible and practically grounded. Early, continuous collaboration fosters shared understanding, reduces ambiguity, and minimises the risk of downstream rework. As one participant noted, \textit{“We sit together—tool developers, annotators, and product owners—before finalising requirements.”} (ID2). This co-creation approach enables teams to anticipate challenges, align expectations, and produce more accurate and efficient annotation workflows.

\begin{tcolorbox}[
enhanced,
boxrule=0.3pt,
colback=white,
colframe=black!50,
left=2pt,
right=2pt,
top=2pt,
bottom=2pt,
sharp corners,
before skip=4pt,
after skip=4pt
]
\textit{$\triangleright$
R2.7 mitigates challenges C2, C3, and C4 by fostering early alignment, promoting shared understanding, and co-creating technically sound annotation requirements.}
\end{tcolorbox}

\textbf{(R2.8) Develop Standardised and Modular Guideline Templates }
Creating standardised, modular templates for data annotation requirements improves consistency, reduces redundancy, and streamlines the development process across projects. These templates serve as adaptable blueprints that maintain core elements like structure, quality benchmarks, and examples, while allowing task-specific customisation. As one participant noted, \textit{“We need more structured templates—not reinventing guidelines for every project.”} (ID15). Templates support quicker onboarding, reduce misinterpretation, and enhance cross-team coordination, ultimately boosting annotation quality and scalability.

\begin{tcolorbox}[
enhanced,
boxrule=0.3pt,
colback=white,
colframe=black!50,
left=2pt,
right=2pt,
top=2pt,
bottom=2pt,
sharp corners,
before skip=4pt,
after skip=4pt
]
\textit{$\triangleright$
R2.8 mitigates challenges C2 , C4, and C5 by enabling consistent reuse, reducing authoring effort, and improving cross-team understanding.}
\end{tcolorbox}

\textbf{(R2.9) Allocate Resources for Structured Requirement Planning and Documentation }
Practitioners stress the need to allocate time, expertise, and tools early in the project to define clear and coherent annotation requirements. Without these resources, teams risk ambiguous  guidelines and costly rework—\textit{“If we skip planning to save time, we end up redoing the work later.”} (ID15). Early investment in annotation requirement development helps avoid midstream confusion and establishes a stable foundation for high-quality annotations.

\begin{tcolorbox}[
enhanced,
boxrule=0.3pt,
colback=white,
colframe=black!50,
left=2pt,
right=2pt,
top=2pt,
bottom=2pt,
sharp corners,
before skip=4pt,
after skip=4pt
]
\textit{$\triangleright$
R2.9 mitigates challenges C3, C4, and C5.3 by ensuring dedicated resources are available for early planning.}
\end{tcolorbox}
\textbf{(R2.10) Integrate Change Management and Version Control }
As annotation requirements evolve due to system changes or emerging edge cases, structured version control is vital to maintain consistency and traceability. Practitioners stress the need for formal processes that log changes, link batches to specific guideline versions, and inform annotators promptly—\textit{“We need proper versioning—so we know which batch followed which version of the rules.”} (ID17). Version control helps avoid outdated instructions, supports auditability, and preserves dataset integrity across development cycles.

\begin{tcolorbox}[
enhanced,
boxrule=0.3pt,
colback=white,
colframe=black!50,
left=2pt,
right=2pt,
top=2pt,
bottom=2pt,
sharp corners,
before skip=4pt,
after skip=4pt
]
\textit{$\triangleright$
R2.10 mitigates challenges C3, C4, and C5.4 by establishing traceability, synchronising changes across annotation efforts, and ensuring that evolving requirements are applied consistently.}
\end{tcolorbox}

\textbf{(R3) Quality Assurance for Data Annotation Requirements }
High-quality data annotation requirements are critical for producing accurate, consistent, and compliant annotations. Practitioners emphasise structured QA mechanisms—such as multi-stage reviews, inter-annotator agreement checks, and feedback loops—to identify and resolve ambiguity early. \textit{“Multiple reviewers verify tasks to eliminate ambiguity and maintain consistency.”} (ID1). Continuous training and automated validation further ensure the robust application of guidelines and data reliability.

\begin{tcolorbox}[
enhanced,
boxrule=0.3pt,
colback=white,
colframe=black!50,
left=2pt,
right=2pt,
top=2pt,
bottom=2pt,
sharp corners,
before skip=4pt,
after skip=4pt
]
\textit{$\triangleright$
R3 mitigates challenges C2, C4, and C5.4 by embedding quality checks, review structures, and validation mechanisms that ensure annotation requirements remain accurate, interpretable, and reliable across teams.}
\end{tcolorbox}

\textbf{(R3.1) Training, Systematic Quality Checks, and Clear Quality Criteria }
Robust annotations rely on well-trained annotators, systematic quality control, and clear evaluation criteria. Domain-specific training and structured feedback—such as expert validation and real-time error detection—equip annotators to handle ambiguity and edge cases. \textit{“Annotators do not know the context unless they’re trained well enough.”} (ID3).

\begin{tcolorbox}[
enhanced,
boxrule=0.3pt,
colback=white,
colframe=black!50,
left=2pt,
right=2pt,
top=2pt,
bottom=2pt,
sharp corners,
before skip=4pt,
after skip=4pt
]
\textit{$\triangleright$
R3.1 mitigates challenges C2, C4, and C5.2 by building annotator expertise, defining clear benchmarks, and enforcing quality validation mechanisms.}
\end{tcolorbox}

\textbf{(R3.2) Manage Consistency Across Annotators }
Consistency is key to dataset reliability. Practices such as consensus-driven guidelines, inter-annotator agreement checks, calibration sessions, and expert reviews help ensure alignment. \textit{“Consensus is essential—annotators following the same data annotation requirements should produce aligned outputs.”} (ID1).

\begin{tcolorbox}[
enhanced,
boxrule=0.3pt,
colback=white,
colframe=black!50,
left=2pt,
right=2pt,
top=2pt,
bottom=2pt,
sharp corners,
before skip=4pt,
after skip=4pt
]
\textit{$\triangleright$
R3.2 mitigates challenges C2,C4, and C5.2 by fostering annotator alignment and expert calibration.}
\end{tcolorbox}

\textbf{(R3.3) Manage Quality to Achieve Scalability }
As projects grow, scalable QA becomes vital. Teams should use granular performance tracking, semi-automated reviews, and targeted retraining to maintain quality without sacrificing speed. \textit{“We track performance at a granular level to identify errors and apply targeted retraining.”} (ID2).

\begin{tcolorbox}[
enhanced,
boxrule=0.3pt,
colback=white,
colframe=black!50,
left=2pt,
right=2pt,
top=2pt,
bottom=2pt,
sharp corners,
before skip=4pt,
after skip=4pt
]
\textit{$\triangleright$
R3.3 mitigates challenges C4, C5.1, and C5.4 by embedding scalable QA mechanisms and maintaining annotation performance.}
\end{tcolorbox}

\textbf{(R3.4) Incentivising Quality Through Performance Structures }
To reduce rushed and error-prone work, teams should reward accuracy and careful handling of complex cases. Quality-based incentives promote accountability and reduce the trade-off between speed and correctness. As mentioned \textit{“We shouldn’t rush edge cases. Maybe quality-based bonuses would help.”} (ID16).

\begin{tcolorbox}[
enhanced,
boxrule=0.3pt,
colback=white,
colframe=black!50,
left=2pt,
right=2pt,
top=2pt,
bottom=2pt,
sharp corners,
before skip=4pt,
after skip=4pt
]
\textit{$\triangleright$
R3.4 mitigates challenges C4, C5.1, and C5.2 by aligning annotator incentives with quality goals and promoting careful edge case handling.}
\end{tcolorbox}

\textbf{(R3.5) Transparency and Traceability in Annotation Workflows }
Traceability is essential for audits, debugging, and accountability. Linking annotations to specific guideline versions and tool configurations ensures reproducibility and regulatory compliance. \textit{“We couldn’t trace which annotation version caused which issue—traceability was missing.”} (ID17).

\begin{tcolorbox}[
enhanced,
boxrule=0.3pt,
colback=white,
colframe=black!50,
left=2pt,
right=2pt,
top=2pt,
bottom=2pt,
sharp corners,
before skip=4pt,
after skip=4pt
]
\textit{$\triangleright$
R3.5 mitigates challenges C4, C5.3, and C5.4 by enabling traceable workflows that support auditing, debugging, and continuous improvement.}
\end{tcolorbox}

\textbf{(R3.6) Align Tool Capabilities with Requirement Complexity }
As annotation requirements grow more complex, tools must adapt to support nuanced rules, context-aware labelling, and multi-step workflows. When tools lag behind evolving needs, they become bottlenecks—\textit{“You change the requirement, but the tool doesn’t support it. That’s a bottleneck.”} (ID12). Misalignment between tools and requirements leads to inconsistency, confusion, and rework. To maintain efficiency and traceability, tools should be flexible, integrate with evolving guidelines, and enable seamless implementation of complex requirements.

\begin{tcolorbox}[
enhanced,
boxrule=0.3pt,
colback=white,
colframe=black!50,
left=2pt,
right=2pt,
top=2pt,
bottom=2pt,
sharp corners,
before skip=4pt,
after skip=4pt
]
\textit{$\triangleright$
R3.6 mitigates challenges C3, C4, and C5.4 by ensuring that tools remain compatible with complex and changing annotation requirements, reducing implementation mismatches and annotation friction.}
\end{tcolorbox}

\begin{figure}[htbp]  
    \centering
    \includegraphics[width=0.5\textwidth]{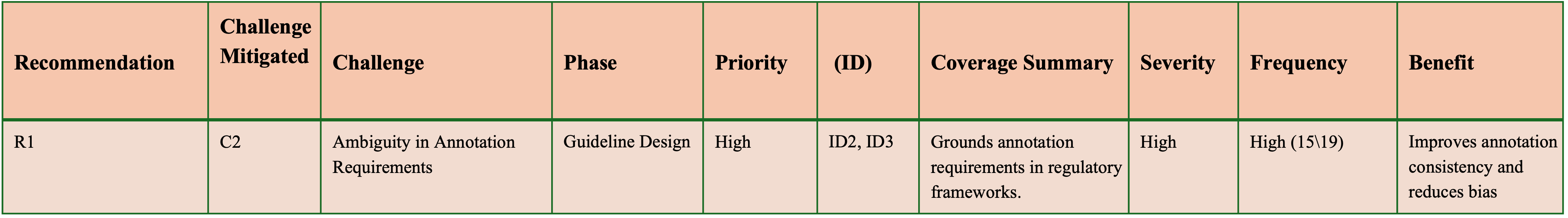} 
    \caption{A snapshot of our publicly available  mapping of practitioner-informed recommendations to identified challenges in AIePS development. The full final repository can be found via \href{https://dataverse.harvard.edu/file.xhtml?fileId=11737897&version=3.0}{\textcolor{blue}{→ link: Practitioners informed recommendations mapped to the identified challenges}}.}
    \label{fig:R2C}
\end{figure}

\textbf{Recommendation Mapping to Challenges Summary} To support traceability, we map each recommendation to the specific challenges it addresses in the AIePS development pipeline, an example illustrated in Fig.~\ref{fig:R2C}.
\vspace{-0.2cm}

The analysis reveals that challenges such as ambiguity in annotation requirements, inconsistent guideline interpretation, and edge case definition are not only highly prevalent (mentioned in over $15$ out of $19 $ interviews) but also critically impactful to system performance. Corresponding recommendations—such as embedding legal compliance (R1.1), defining safety-centric requirements (R1.2), and simplifying guidelines with iterative feedback (R2.1, R2.3)—were consistently rated as high priority and highly beneficial . Edge case-related challenges (e.g., C1, C1.2) stand out with the highest criticality scores, indicating their potential to cause safety failures if left unaddressed. Many recommendations (e.g., R2.5, R3.2) target cross-cutting issues like resource limitations and workforce scalability, offering benefits such as improved throughput, reduced rework, and enhanced model generalisation. Our traceable mapping of recommendations to challenges—enriched with metadata like annotation phase, empirical support, and expected benefit—provides a practical foundation for improving annotation pipelines in safety-critical AI domains.

\vspace{-0.2cm}

\begin{tcolorbox}[
  enhanced,
  colback=gray!5!white,      
  colframe=gray!60!black,    
  coltitle=white,            
  title=\textbf{Key Takeaways from RQ2},
  fonttitle=\bfseries\large,
  sharp corners,
  boxrule=0.6pt,
  drop shadow southeast,
  left=0.5mm,                
  right=2mm, top=1mm, bottom=1mm,
  before skip=10pt, after skip=10pt,
  colbacktitle=gray!60!black 
  ]

\vspace{0.3em}

\begin{itemize}

    \item Iterative feedback loops (plan–do–check–act) improve annotation quality and guideline maturity.
    \item Edge cases require proactive strategies, expert validation, and synthetic augmentation.
    \item Automation can enhance scalability and consistency, but human oversight remains essential.
    \item Cross-functional collaboration ensures guidelines are technically feasible and contextually grounded.

\end{itemize}
\end{tcolorbox}

\begin{figure*}[htbp]  
    \centering
    \includegraphics[width=1.0\textwidth]{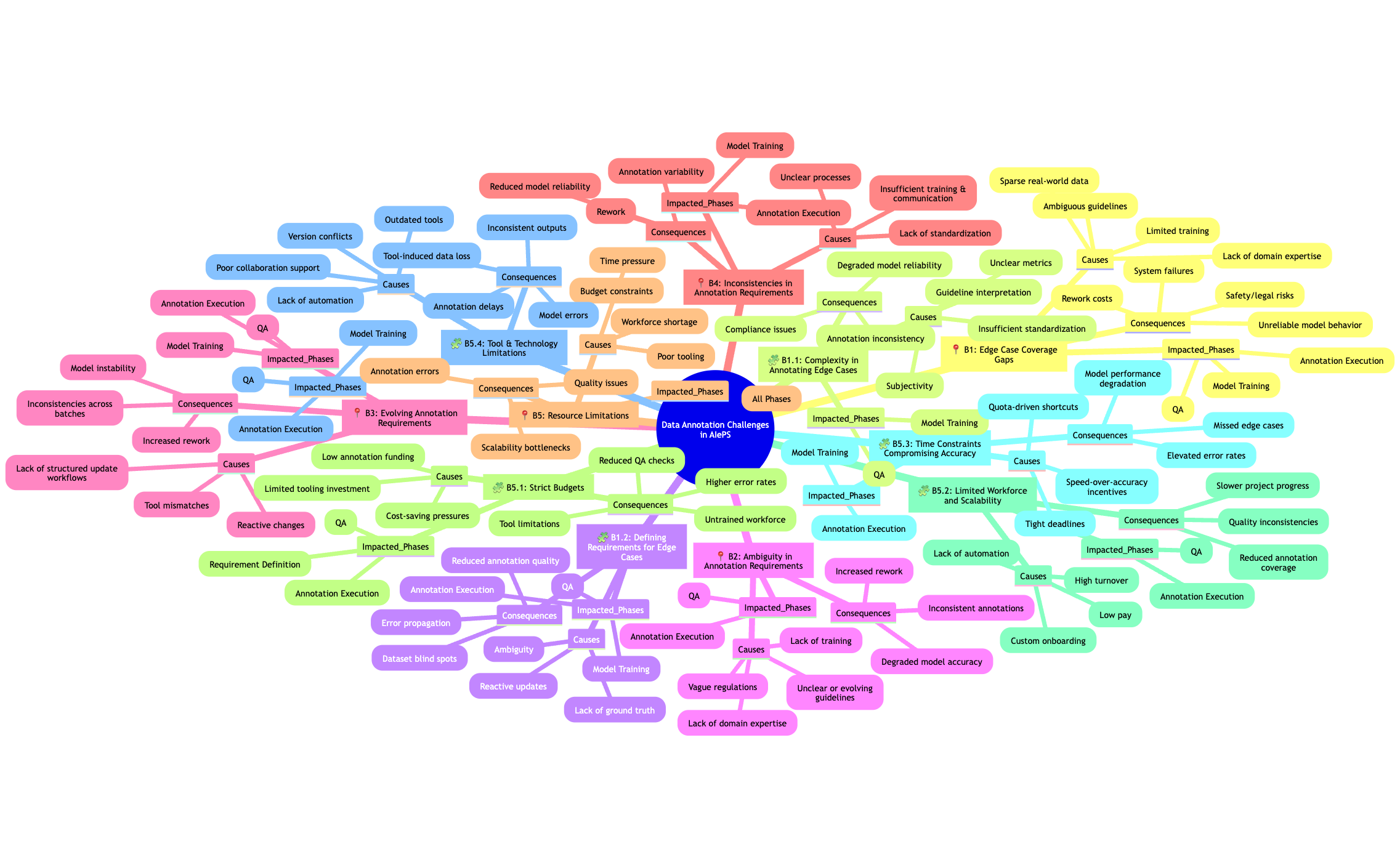} 
    \caption{ Causal mind map of data annotation challenges in AI-enabled perception systems (AIePS).  }
   \label{fig:mindmap}
\end{figure*}
\section{Threats to validity}

We consider \emph{\textbf{construction validity}} as the aim of the study is to explore a real industrial challenge; it is crucial to identify experts in the relevant industries. Hence, we identified the profiles of experts and their companies, and selected participants from the pool of experts who matched these profiles.
In addition, to ensure the effectiveness of the interview, the interview protocol was subjected to an internal review by the other researchers in the team and evaluated through two pilot studies. For \emph{\textbf{internal validity}}
As the results of the study are highly dependent on the opinions of the participants, their responses may reflect personal bias. To mitigate this, participants were selected from six companies with diverse backgrounds and different roles in the distributed development of AIePS.
Five researchers from academia, involved as pioneers or chief investigators in AI development, are also included in the study to enhance diversity.  
While ensuring \emph{\textbf{external validity}} is crucial to ensure the study's results are generalizable, it is also essential to emphasise diversity. In this study, participants were selected from two European countries and the UK. However, due to standardised processes  (e.g., ISO/IEC 5259, IEEE P2801, ISO 26262, SAE J3016)  and the international market of companies involved in the development of AIePS, there is a high degree of consensus on the process to meet the requirements in different regions.  For transparency and reproducibility, we have shared our replication materials.

The \emph{conclusion validity} of this study is supported through systematic data collection and analysis steps. Semi-structured interviews with $19$ interviews ($20$ participants)  from both industry and academia ensured diverse perspectives. Thematic analysis followed a mixed coding approach, applying both deductive and inductive codes. To enhance coding accuracy, at least two researchers independently coded the initial interviews, with iterative peer reviews and discussions resolving discrepancies, achieving high inter-coder agreement (Cohen’s Kappa $= 0.8$). The collaborative involvement of five researchers further strengthened theme extraction and reduced individual bias. Identified themes and recommendations were cross-validated with existing literature, ensuring consistency and alignment with prior research. Collectively, these steps reinforce the logical connection between data and conclusions.

\section{Discussion, Reflections, and Conclusions}

\textbf{Repositioning Data Annotation Requirements as Formal RE Artefacts } Prior RE studies in safety-critical AI systems have largely focused on high-level system safety, traceability, and stakeholder alignment \cite{habibullah2023requirements,heyn2023automotive,dey2023multi}. Still, they overlook the specific role of data annotation requirements. Our study uniquely treats these requirements as formal RE artefacts and shifts attention to the pre-training data pipeline, where ambiguities and inconsistencies can directly affect AIePS performance. Unlike previous conceptual frameworks \cite{dey2023multi} or ecosystem-level analyses \cite{heyn2023automotive}, we provide an empirically grounded, fine-grained classification of annotation challenges (C1–C5.4), their severity, and a mapping to actionable recommendations. Whereas prior work has highlighted the quality of annotations \cite{mohammedali2023influence,beck2023quality}, we take it a step further by embedding annotation requirements within structured RE processes, extending principles such as traceability and refinement \cite{IREB2024} to the data-centric development of safe AI systems.
In current industrial practice, data annotation requirements are typically defined by domain experts such as safety specialists and perception engineers, rather than by traditional system analysts or requirements engineers. This reflects the highly domain-specific nature of annotation tasks, which require deep expertise in traffic scenarios, sensor data, and edge-case behaviour. Our study highlights how RE principles can complement this process, thereby introducing structure, traceability, and formalisation into a space where RE professionals have not yet been systematically involved.

\textbf{Evolving Practices and System-Level Impacts} Annotation requirements in AIePS evolve through an iterative, feedback-driven, and stakeholder-centric process rather than being predefined—echoing findings by Habibullah et al. \cite{habibullah2023requirements} and Heyn et al. \cite{heyn2023automotive}. This dynamic evolution is reinforced by Watson et al. \cite{watson2023augmented} and Klie et al. \cite{klie2024analyzing}, while Dey and Lee \cite{dey2023multi} emphasise the involvement of multiple stakeholders, including clients, annotators, and engineers. These practices align with IREB (2024) \cite{IREB2024}  principles of iterative refinement, collaboration, and traceability. Yet, unlike conventional RE artefacts, annotation requirements often lack formal mechanisms for validation, impact analysis, and ownership, leading to inconsistencies. Their evolution is also shaped by model feedback, epistemic uncertainty, and operational constraints, challenging traditional goal-driven elicitation approaches.

\textbf{RQ1:} Our study identifies key challenges in ensuring high-quality annotation requirements for AIePS, especially in autonomous driving. While prior work has emphasised uncommon scenarios \cite{bachute2021autonomous, ren2023challenges}, we show that inter-annotator disagreement and evolving requirements further hinder edge case coverage and introduce inconsistencies. In contrast to studies on automation \cite{demrozi2021towards} or scenario completeness \cite{ren2023challenges}, our findings reveal industry-specific issues such as persistent ambiguity due to reactive updates, insufficient documentation, and weak feedback loops—even in the presence of annotation standards \cite{khattak2021taxonomy}. These lead to operational burdens like re-annotation, communication gaps, and inconsistent interpretations. Moreover, resource limitations—such as time, staffing, and tooling—significantly affect how annotation requirements are defined and applied. High turnover and constrained budgets compromise training quality, leading to shortcuts that particularly impact edge cases. Our study empirically demonstrates how these factors degrade annotation consistency and, ultimately, AIePS reliability as presented in detail in Fig.~\ref{fig:mindmap}.

\textbf{RQ2:} Practitioners address annotation ambiguity and inconsistency using structured quality assurance mechanisms such as independent reviews and inter-annotator agreement checks. Given that guidelines often exceed $140$ pages \cite{sharma2020evaluation}, simplification, iterative feedback, visual tools, and embedded QA workflows are essential to reduce errors and improve clarity. Cross-functional collaboration among data scientists, coordinators, and domain experts aligns annotation practices with system goals, as also supported by Heyn et al. \cite{heyn2023automotive}. While automation can support routine tasks, human oversight remains critical for handling edge cases and complex scenarios \cite{watson2023augmented}. Synthetic data and simulation help supplement rare cases \cite{bachute2021autonomous}, while bias mitigation and compliance with GDPR and the AI Act are foundational \cite{habibullah2023requirements, yang2023uncertainties}. Scalability requires embedded training, monitoring, and modular QA frameworks \cite{beck2023quality}. Unified standards can improve consistency and reduce onboarding time \cite{klie2024analyzing}, yet siloed teams, a lack of shared ownership, and limited tooling often hinder collaborative implementation. Our study reinforces that data annotation requirements have a profound influence on annotation processes, dataset quality, and AIePS performance. The causal effect analysis of data annotation requirements in the AIePS development pipeline is detailed in Fig.~\ref{fig:mindmap}. Clear, specific requirements enhance consistency and reduce errors \cite{beck2023quality}, while vague ones lead to variability and mislabeling \cite{nassar2019assessing, klie2024analyzing}. Contrary to assumptions that annotation precision directly ensures AIePS accuracy \cite{yaqoob2019autonomous, bachute2021autonomous}, our findings stress the importance of edge case representation and feedback loops \cite{wang2024survey}, showing that systematic mislabeling can be more detrimental than random errors \cite{borg2018safely}.
The recommendations presented in this study are not intended as a fixed sequence but rather as a toolbox of targeted practices. Each recommendation is explicitly mapped to the challenge(s) it mitigates, allowing practitioners to adaptively prioritise those linked to the most critical issues in their specific context. For example, projects facing ambiguity can emphasise clarity-enhancing measures, whereas those constrained by resources may prioritise efficiency-oriented practices.

Our recommendations should be understood as practical enablers for applying existing standards in day-to-day annotation practice. For instance, clarity-enhancing practices. directly operationalise ISO/IEC 5259’s requirements on data clarity and consistency, as well as traceability-oriented recommendations. support ISO 26262’s safety assurance principles and efficiency-focused measures, complementing IEEE P2801’s lifecycle management of AI systems. In this way, our work bridges the gap between what is specified in international standards and the how of implementing these requirements in real-world AI-enabled perception system development.

\vspace{-3pt}
\textbf{Reflection and Limitations} while the recommendations offer practical guidance, challenges remain—especially the trade-offs between accuracy and scalability in resource-limited settings. Automation aids efficiency but can introduce bias without expert oversight. Collaboration is often undermined by siloed teams, unclear roles, and inadequate tooling. These issues underscore the need for integrated, traceable, and role-aware RE for AI. Future research should investigate how requirements evolve over time in real-world contexts, compare practices across domains, and develop tools to manage change, ensure traceability, and provide automated feedback.

\textbf{Implications for Research and Practices} 
 Our study highlights significant research and practice gaps in managing annotation requirements for AIePS, including the lack of formal quality metrics and a limited understanding of their impact on data and system safety. It calls for integrating RE practices—such as change analysis and stakeholder alignment—into annotation workflows. Socio-technical factors, such as team coordination, tooling, and legal/ethical implementation, remain underexplored and require theory-driven, empirical research. For practitioners, the study recommends treating annotation requirements as structured, version-controlled artefacts embedded in RE processes. Clear guidelines, iterative feedback, expert input for edge cases, and tool-based support can improve clarity and traceability. Automation must be balanced with human oversight. Regulatory compliance, annotator training, and quality assurance are essential, with cross-functional collaboration key to developing high-quality, context-aware annotation requirements in safety-critical AI systems.


\bibliographystyle{plain} 
\bibliography{bib} 

\end{document}